\documentstyle[prl,twocolum,aps]{revtex}
\newcommand{\qq}{\mbox{\boldmath$q$}}
\newcommand{\FF}{\mbox{\boldmath$F$}}

\begin{document}
\draft
\title{Supersymmetry solution for finitely extensible dumbbell model}

\author{P.\ Ilg \thanks{E-mail: pilg@ifp.mat.ethz.ch}, I.\ V.\ Karlin\\
{\it Department of Materials, Institute of Polymers, ETH Z\"{u}rich, 
	CH--8092 Z\"{u}rich, Switzerland}\\
	S.\ Succi\\
{\it Institute Applied Computing, CNR, Viale Policlinico 137, 00161 Rome, 
	Italy}}
\maketitle

\begin{abstract}
Exact relaxation times and eigenfunctions for a simple mechanical 
model of polymer dynamics are obtained using supersymmetry methods 
of quantum mechanics. 
The model includes the finite extensibility of the 
molecule and does not make use of the 
self-consistently averaging approximation.
The finite extensibility reduces the relaxation times when compared
to a linear force. 
The linear viscoelastic behaviour is obtained in the form of the 
``generalized Maxwell model''. 
Using these results, a numerical integration scheme is proposed in the 
presence of a given flow kinematics.
\end{abstract}

\section{Introduction}
Simple mechanical models are helpful for  
understanding the complex dynamical behaviour of polymer molecules. 
The so-called elastic dumbbell constitutes the simplest model 
that captures the effects of stretching and orientation of the polymer 
\cite{Bird87}.
In this model the polymer is represented by two beads which are connected 
by a spring. 
The classical Hookean dumbbell model \cite{Bird87} is 
characterised by a linear spring force. 
Its relaxation times decrease like the inverse of the mode number
and the relaxation modulus is of a single exponential form. 
As noted by many authors, the finite extensibility of the molecules 
has to be included in order to avoid unphysical behaviour such as infinite 
viscosities predicted by the Hookean model.
It is even argued 
``that taking into account the nonlinear stretching of the polymer 
molecules is the most important correction that should be made to 
the Hookean model in order to describe real systems''
(Ref. \cite{Bird87}, Chap. 13, p. 81).
Molecular arguments suggest the finitely extensible spring force law 
to be given by the inverse Langevin function \cite{Tr75}. 
Very often, rational \cite{Wa72} or Pad\'e approximations \cite{Co91} 
to the inverse Langevin function are used since they are analytically 
more tractable. 
All these approximations show the desired physical characteristics: 
a linear regime for small extensions of the spring and a divergence at a 
finite extension. These forces therefore prohibit the 
molecule to stretch beyond a certain value which is related to the 
total length of the chain. 

In general, the finite extensibility of the molecules prohibits 
analytical expressions for the spectrum of relaxation times and the 
relaxation modulus. 
The shear relaxation modulus for the Warner spring force 
(so-called FENE model, \cite{Bird87,Wa72}) has been determined 
numerically \cite{HO97}.
The results are very close to those obtained for the Peterlin 
approximation to the Warner force (FENE-P model, \cite{Bird87,BDJ80}). 
The effect of the finite extensibility on the spectrum of relaxation 
times remains less studied. One might speculate that the finite 
extensibility decreases the slowest relaxation time as described 
by the Peterlin approximation. However, knowledge of the whole 
spectrum of relaxation times would be very desirable for understanding 
the dynamics of dumbbell models. It would allow, for example, to 
identify the number of relevant slow modes that contribute to 
the polymer contribution to the stress. 
Work in this direction was performed in \cite{ZKD00} where the 
universal constitutive equation of dilute polymeric solutions was 
found. There, it was assumed that in the limit of low Deborah and 
Weissenberg number the modes corresponding to the two lowest 
eigenvalues dominate. 

Here, we compute the relaxation times, eigenfunctions and the 
relaxation modulus of a finitely extensible dumbbell model exactly. 
The model is a one-dimensional version of the finitely extensible 
dumbbell models that uses trigonometric functions to approximate 
the inverse Langevin force. It is demonstrated that this force 
approximates the inverse Langevin function as good or even better 
than the Warner force frequently used in Brownian dynamics simulations. 
The present study therefore clarifies the effect of the finitely 
extensible spring force on the spectrum of relaxation times, 
eigenfunctions and the relaxation modulus.

\section{Model description}
Let $f(\qq;t)$ denote the distribution function of the connector vector 
$\qq$ of the dumbbell at time $t$. The time evolution of $f$ is 
given by the well-known Fokker-Planck equation \cite{Bird87}
\begin{equation} \label{FPeq}
	\partial_t f = D\nabla \left\{ [\nabla U] + \nabla \right\} f 
	\equiv L_{\rm FP} f
\end{equation}
where we have assumed that the spring force $\FF$ can be derived from the 
dimensionless potential $U$ via $\FF(\qq)=k_{\rm B}T\nabla U(\qq)$.
The diffusion constant $D=2k_{\rm B}T/\zeta$ is assumed to be independent 
of $\qq$ which means we neglect hydrodynamic interactions. 
Note that the equilibrium distribution $f_{\rm eq}$ is given by 
$f_{\rm eq} = N_{\rm eq} e^{-U}$, $N_{\rm eq}$ being a normalization 
constant. 
As is well-known the Fokker-Planck equation (\ref{FPeq}) can be transformed 
into an equivalent imaginary time Schr\"odinger equation \cite{Cu91,Ris96}. 
If $\psi$ is a solution to 
\begin{equation} \label{Schreq}
	-\partial_t \psi = DH \psi
\end{equation}
with the operator $H = -(1/D)e^{U/2}L_{\rm FP}e^{-U/2}$, then the function 
$f(\qq;t)=\psi(\qq;t)f_{\rm eq}^{1/2}(\qq)$ 
solves the original Fokker-Planck equation (\ref{FPeq}). 
In particular, solving the eigenvalue problem 
$H\psi_n=\epsilon_n\psi_n$ 
leads immediately to the solution of the eigenvalue problem of the original 
Fokker-Planck operator $L_{\rm FP}f_n=-\lambda_n^{-1}f_n$ with 
$\lambda_n^{-1}=D\epsilon_n$ and $f_n=\psi_nf_{\rm eq}^{1/2}$, where 
factors $D$ are introduced for later convenience.  
Inserting (\ref{FPeq}) for the Fokker-Planck operator 
$H$ yields the form of the Schr\"odinger operator
\begin{equation} \label{Schop}
	H = -\nabla^2 + V(\qq),
\end{equation}
where the Schr\"odinger potential $V$ is given by $V=\Phi^2 - \nabla \Phi$, 
with $\Phi=\nabla U/2$. 
Therefore, each solvable potential for the Schr\"odinger equation serves 
as a solvable model for the Fokker-Planck equation. 
In the following, we benefit from the fact that simple Schr\"odinger 
potentials $V$ can have quite complicated counterparts $U$ for the 
Fokker-Planck equation. 
Note also that operator $H$ is Hermitian which is in general not the 
case for $L_{\rm FP}$. 

The simplest exactly solvable potential for the Schr\"odinger equation 
related to finite extensibility is the three dimensional 
spherical symmetric infinite well potential.  
Up to a constant, the corresponding potential $U$ of the Fokker-Planck 
equation is given by $U=-2\ln \psi_{\rm eq}$, where $\psi_{\rm eq}$ is 
the ground state of the Schr\"odinger problem. 
In the present case, the ground state is known to be 
$\psi_{\rm eq} \sim \sin(\pi q/q_0)/q$ with $q=|\qq|$  
leading to the force law
\begin{equation} \label{3Dforce}
	\FF(\qq) = 2k_{\rm B}T 
	\frac{1-(\pi q/q_0)\cot(\pi q/q_0)}{q}\frac{\qq}{q}. 
\end{equation}
Since the force (\ref{3Dforce}) is one-parametric, the ``spring constant'' 
is no adjustable parameter but determined by the maximum extension $q_0$. 
If we define the spring constant $h$ in the linear regime of 
(\ref{3Dforce}), the dimensionless finite extensibility parameter 
$b=hq_0^2/(k_{\rm B}T)$ is fixed: $b=b_3=2\pi^2/3$. 
Parameter $b$ is known to denote roughly the number of monomer units 
represented by the dumbbell and should therefore be a large number. 
Thus, (\ref{3Dforce}) is of limited use. 
However, if $b=b_3$ is acceptable the force (\ref{3Dforce}) compares very 
well to the inverse Langevin force and Cohen's Pad\'e approximation, 
as shown in the inset of Fig. \ref{forces}.  
In this case, exact eigenfunctions of the Schr\"odinger operator are known 
to be the spherical Bessel functions and the exact eigenvalues are given by 
the zeros of this functions.

Let us now consider the one-dimensional case. 
One-dimensional nonlinear dumbbell models serve as toy models for various 
approximations that have been proposed to obtain closed constitutive 
equations for polymer solutions \cite{Keu98}. 
In addition, the one-dimensional models are believed to describe the 
elongational behaviour of the polymer molecule reasonably \cite{Keu97}. 
The one-dimensional version of the operators $L_{\rm FP}$ and $H$ is 
obtained by replacing $\qq$ with $q$ and $\nabla$ with $d/dq$.  

To account for the finite extensibility we propose the following two 
parameter family of force laws
\begin{equation} \label{1Dforce}
	F(q)=hq_1\tan(q/q_1),\quad \mbox{for}\ -q_0 < q < q_0
\end{equation}
where $q_1=2q_0/\pi$ is determined by the maximum extension $q_0$ and 
$h$ denotes the ``spring constant'' in the sense $F(q) \to hq$ for 
$q\to 0$.  
In Fig. \ref{forces} we plot the force $F/h$ as a function of the 
reduced extension $q/q_0$. 
Fig. \ref{forces} also shows the Warner force \cite{Wa72} 
$F(q)=hq/(1-q^2/q_0^2)$, 
the inverse Langevin force $(h/3)L^{-1}(q/q_0)$, 
with $L(x)=\coth(x)-1/x$ and Cohen's Pad\'e approximation 
$F(q)=(hq/3)(3-q^2/q_0^2)/(1-q^2/q_0^2)$ \cite{Co91}. 
It is easily verified that these approximations 
give the correct limiting behaviour not only for small but 
also near the maximum extension. 
In addition, as can be seen from Fig. \ref{forces}, the force 
(\ref{1Dforce}) serves 
as an even better approximation to the inverse Langevin force as does the 
Warner force. 
Therefore we accept (\ref{1Dforce}) as a reasonable force law to 
substitute for the inverse Langevin force in one dimension. 

The potential $U$ from which the force (\ref{1Dforce}) can be derived is 
$U(q)=-b_1\ln\cos(q/q_1)$, where we introduced the dimensionless 
parameter $b_1=hq_1^2/(k_{\rm B}T)=(2/\pi)^2b$. 
The equilibrium distribution function of the Fokker-Planck equation is 
$f_{\rm eq}(q) = N_{\rm eq} \cos^{b_1}(q/q_1)$, with 
$N_{\rm eq}^{-1} = q_12^{b_1}B[(b_1+1)/2,(b_1+1)/2]$, where 
$B[x,y]$ is the Beta function.  
The equilibrium distribution is very close to the one corresponding to 
the Warner force
$f_{\rm eq}(q) = N_{\rm eq} (1-q^2/q_0^2)^{b/2}$, with 
$N_{\rm eq}^{-1} = q_0B[1/2,(b+2)/2]$. 

\section{Supersymmetry solution}
To obtain exact eigenvalues and eigenfunctions of the Fokker-Planck 
equation (\ref{FPeq}) we exploit supersymmetry methods 
for the corresponding Schr\"odinger problem. 
Note, that the model force (\ref{1Dforce}) belongs to the class of 
so-called Poeschl-Teller potentials which may also be solved exactly by 
Schr\"odinger's factorization method \cite{Ju96}.
Consider the original potential $U$ together with the inverted 
potential $-U$ and denote by $f_-$ the solution for the potential $U$ 
and $f_+$ for $-U$. 
As done before, defining 
$f_\pm=\sqrt{N_{\rm eq}}e^{\pm U/2}\psi_\pm$ we arrive 
at $-(1/D)\partial_t \psi_\pm = H_\pm \psi_\pm$ with operators
$H_\pm = \nabla^2 + V_\pm(q)$. The ``partner potentials'' are given 
by $V_\pm(q)=\Phi^2(q) \pm \Phi'(q)$ with $\Phi=U'/2$, where 
$\Phi'$ and $U'$ denote the derivative of $\Phi$ resp. $U$ with respect 
to $q$.  
Schr\"odinger operators $H_\pm$ are very well studied in the literature 
since they occur in the so-called 
``Witten model'' \cite{Wi}, which is the simplest model that shows all 
typical features of supersymmetric quantum mechanics. 

Rewriting the force (\ref{1Dforce}) as $F=2k_{\rm B}T\Phi$ we obtain 
$\Phi_\alpha(q)=\alpha\tan(q/q_1)$, for $-q_0 < q < q_0$. 
We denote explicitly the dependence on the parameter 
$\alpha=b_1/(2q_1)$ which is proportional to the spring constant $h$. 
To proceed further we take advantage of the concept of ``shape invariance'' 
introduced by Gendenshte\^{\i}n \cite{Ge83}. 
The partner potentials 
$V_\pm(\alpha,q)=\Phi_\alpha^2(q)\pm \Phi'_\alpha(q)$ 
are called shape invariant if they are related by 
$V_+(\alpha_k,q)=V_-(\alpha_{k+1},q) + R(\alpha_{k+1})$ where the new 
parameter $\alpha_{k+1}$ is a function of $\alpha_k$. 
In our case, $\Phi_{\alpha}=\alpha\tan(q/q_1)$, shape invariance is easily 
verified for 
$\alpha_{k+1}=\alpha_k+1/q_1$, $\alpha_0=\alpha$ and 
$R(\alpha_{k+1})=\alpha^2_{k+1}-\alpha^2_k$. 
For shape invariant potentials it 
is possible to define a series of operators that are isospectral except 
for the lowest eigenvalue which is $\sum_{k=1}^n R(\alpha_k)$. 
Going back to the original Schr\"odinger operator with lowest eigenvalue 
zero the complete spectrum is found to be 
$\epsilon_n=\sum_{k=1}^nR(\alpha_k)$ \cite{Ju96,Ge83}. 
Inserting the special form of $R$ and multiplying by $D$,  
the exact (inverse) eigenvalues of the Fokker-Planck operator (\ref{FPeq}) 
are given by 
\begin{equation} \label{1Dvalues}
	\lambda_n = 2\lambda_{\rm H} \left[n+n^2/b_1 \right]^{-1}, \quad 
	n=1,2,\ldots 
\end{equation}
where $\lambda_{\rm H}=\zeta/(4h)$ is the time constant of the Hookean 
dumbbell. 
Note, that the eigenvalue zero corresponds to the equilibrium distribution.
Corresponding eigenfunctions of operator $H$ are 
\begin{equation} \label{1Deigen}
	\psi_n(q) = C_na^\dag_0a^\dag_1\ldots a^\dag_{n-1} 
	\cos^{n+b_1/2}(q/q_1)
\end{equation}
where $a^\dag_k=-d/dq+\Phi_{\alpha_k}$ are generalized creation operators 
and $C_n$ denote normalization constants \cite{Ju96}.
Remember that eigenfunctions of the original Fokker-Planck operator are 
obtained by $f_n=\psi_nf_{\rm eq}^{1/2}$. 
The first eigenfunctions read: $f_0=f_{\rm eq}$, 
$f_1=N_1\sin(q/q_1)f_{\rm eq}$, 
$f_2=N_2[ (b_1+1)\sin^2(q/q_1)-\cos^2(q/q_1) ] f_{\rm eq}$ 
with 
$N_1^2=b_1+2$ and $N_2^2=(b_1+4)/[2(b_1+1)]$.  
Note, that the eigenfunctions are orthonormal, 
$\langle f_n, f_m \rangle = \delta_{nm}$, in the scalar product

\begin{equation} \label{scalar}
	\langle g, h \rangle = \int_{-q_0}^{q_0} 
	f_{\rm eq}^{-1}(q)g(q)h(q)dq. 
\end{equation}
The linear viscoelastic behaviour can be obtained from linear response 
theory. 
For a given flow kinematics the perturbation of the Fokker-Planck 
operator is given by 
$L_{\rm ext}(t)=-\partial_q \kappa(t)q$ where $\kappa(t)$ 
denotes the velocity gradient. 

The polymer contribution to the stress, 
$\tau/(k_{\rm B}T)=\langle \sigma,f \rangle - 1$, $\sigma=f_{\rm eq}qU'$, 
is found to be 
\begin{equation} 
	\tau/(k_{\rm B}T) = \int_{-\infty}^tG(t-t')\kappa(t')dt',
\end{equation}
where the relaxation modulus $G(t)$ is given by the equilibrium 
correlation function 
$G(t)= \langle \sigma,e^{L_{\rm FP}t}\sigma\rangle - 1$. 
Inserting the completeness relation of the eigenfunctions, 
$\delta(q-q')=\sum_n \psi_n(q)\psi_n(q')=
f_{\rm eq}^{-1}(q')\sum_nf_n(q)f_n(q')$, we obtain 
\begin{equation} \label{Maxwell}
	G(t) = \sum_{n=1}^\infty \left| \langle \sigma,f_{2n} \rangle 
	\right|^2 e^{-t/\lambda_{2n}},
\end{equation}
which is of the form of the ``generalized Maxwell model''. 
Note, that only even terms are included in (\ref{Maxwell}) since they 
correspond to symmetric eigenfunctions whereas odd terms, 
corresponding to antisymmetric eigenfunctions, vanish by symmetry. 
Analytical expressions for $G(t)$ resulting from dumbbell models are 
known for linear springs (Hookean dumbbell) where 
$G(t)=e^{-t/\lambda_{\rm H}}$. 
For the Warner spring force no such expression is available. The 
Peterlin approximation \cite{Bird87} leads 
also to a single-exponential form with relaxation time 
$\lambda_{\rm H}[1+1/b]^{-2}$. 
The relaxation modulus (\ref{Maxwell}) in our model shows a 
spectrum of relaxation times $\lambda_n$, 
the lowest one being $\lambda_2=\lambda_{\rm H}[1+2/b_1]^{-1}$. 
The finite extensibility of the dumbbell therefore reduces the longest 
relaxation time when compared to the Hookean dumbbell.  
Only in the limit $b\to \infty$ the result for the Hookean dumbbell 
is recovered. 
As can be seen from Fig. \ref{reltimes}, $G(t)$ is dominated by the lowest 
relaxation time since the relative weight of the higher modes decreases 
rapidly. 
For $b=20$, the relative weight of the first two modes is $99\%$, for 
$b=50$ the first mode carries $99\%$ and for $b\to\infty$ all weights 
except for the first vanish, reflecting the single-exponential form 
of the Hookean dumbbell.  

The relaxation modulus determines the zero-elongational viscosity 
$\bar \eta_0=\int_0^\infty G(t)dt$. As shown in \cite{Bird87}, 
$\bar \eta_0$ can be expressed in terms of second moments of the 
equilibrium distribution function for arbitrary dumbbell models. 
In our model, the resulting integral cannot be done analytically. 
 
Knowledge of the exact relaxation times and eigenfunctions 
(\ref{1Dvalues}) and (\ref{1Deigen}) can further be used to integrate 
the kinetic equation (\ref{FPeq}) in the presence of a given flow field. 
Expanding the distribution function into eigenfunctions, 
$f(q;t)=\sum_n c_n(t)f_n(q)$, leads to 
\begin{equation} \label{coeff}
	\dot c_n = -\lambda_n^{-1} c_n - \kappa(t) \sum_k A_{nk}c_k, 
\end{equation}
with 
$A_{nk}=\langle f_n, \partial_q q f_k \rangle$ 
a constant matrix independent of the flow. 
Coefficients $c_n(t)$ are defined by $c_n(t)=\langle f_n, f(q;t) \rangle$. 
Eqs.\ (\ref{coeff}) are an equivalent formulation of the kinetic 
equation (\ref{FPeq}). 
Conservation of total probability is guaranteed by $\dot c_0 = 0$, which 
is easily verified from (\ref{coeff}). 
The stress $\tau$ is determined by the coefficients $c_n$ via 
$\tau/(k_{\rm B}T)=\sum_{n=1}^\infty \langle \sigma,f_{2n} \rangle c_{2n}$. 
Because of symmetry, only even terms contribute to $\tau$. 
Moreover, Eqs.\ (\ref{coeff}) for even coefficients decouple from those 
for odd coefficients since $A_{nk}$ is nonzero only if $n$ and $k$ are 
both even or both odd. 
Therefore, if we are interested only in the polymer contribution to the 
stress, only even coefficients have to be included in (\ref{coeff}). 
Given $A_{nk}$, Eqs.\ (\ref{coeff}) can be integrated for a 
finite set of eigenfunctions quite efficiently using standard integrators 
for ordinary differential equations. 
How many eigenfunctions have to be included in order to obtain a desired 
accuracy is a delicate question. 
A simple estimation is the following. 
To a first approximation, coefficients $c_n$ are still of exponential 
form but with a different time constant $\lambda_n^{-1}+\kappa A_{nn}$. 
Modes that decrease quadratically with number $n$ will then be negligible. 
This leads to a minimal number of modes $n_c$ that satisfies 
$n_c^2/b_1=n_c+2\lambda_{\rm H}|\kappa(t)A_{n_cn_c}|$.   
If this number is not too large, method (\ref{coeff}) 
is much more efficient than Brownian dynamics simulation.

\section{Conclusions}
We have obtained the spectrum of relaxation times, the eigenfunctions 
and the relaxation modulus for a one-dimensional finitely 
extensible dumbbell model exactly. 
It is observed that the relaxation times are reduced when compared to 
the Hookean dumbbell model. Moreover, the decrease of the relaxation 
times with mode number $n$ shows a crossover from $n^{-1}$ to 
$n^{-2}$ at the value of the finite extensibility parameter $b$. 
The relaxation modulus is obtained in the form of the 
``generalized Maxwell model''. For rather large values of the finite
extensibility parameter $b$, which are commonly believed to be physically 
meaningful (see, e.g. \cite{HO97}), the linear viscoelastic behaviour 
is well described by a single relaxation time. This is because the 
relative weight of the higher modes is decreasing rapidly with 
increasing $b$. 
Therefore, it is not surprising that the linear viscoelastic regime 
is well described by the Peterlin approximation as found in \cite{HO97}. 
To calculate the non-linear dynamical response to a given flow kinematics 
we propose a numerical integration scheme using the expansion of the 
distribution function into the exact eigenfunctions in the absence of flow.

All these results are limited to the one-dimensional case. 
In three dimensions the class of solvable potentials is much more 
restricted. In particular we did not find any relevant two parameter 
potential that can be solved exactly. 
However, there is some evidence, that 
``a simple one-dimensional version of the FENE theory captures 
qualitatively and quantitatively the elongational behaviour of the 
actual three-dimensional theory'' (Ref. \cite{Keu98}, Introduction).

\section*{Acknowledgment}
The authors gratefully acknowledge valuable discussions with H.\ C.\ 
\"Ottinger.


\begin{figure}
\caption{Nonlinear, finitely extensible dumbbell forces are shown as a 
function of the reduced extension. 
Long dashed: Warner force, dotted: the force (\ref{1Dforce}) proposed here, 
dashed: Cohen's Pad\'e approximation, solid line: 
the inverse Langevin function and dot-dashed: the Hookean spring. 
As easily seen from the figure the propositon (\ref{1Dforce}) approximates 
the inverse Langevin function even better than does the most frequently used 
Warner approximation.
The inset shows these forces for the special case $b=b_3$ in three 
dimensions. The same symbols are used as before, but the force  
(\ref{1Dforce}) is replaced by (\ref{3Dforce}).} 
\label{forces}
\end{figure}

\begin{figure}
\caption{The first ten relaxation times $\lambda_n$, $n$ even, 
are displayed as full circles (\ref{1Dvalues}). 
The histogram shows their relative weights 
$|\langle \sigma, f_n \rangle|^2/G(0)$. The dimensionless 
parameter $b$ was chosen to be $b=20$. With increasing $b$ the relaxation 
times approach $1/n$ while $\lambda_2$ accumulates the total weight.}
\label{reltimes}
\end{figure}


\begin{references}

\bibitem{Bird87}R.\ B.\ Bird, C.\ F.\ Curtiss, R.\ C.\ Armstrong, 
O.\ Hassager, {\it Dynamics of Polymeric Liquids, Vol.\ 2, Kinetic Theory}, 
Wiley, New York, 2nd Ed., 1987.
\bibitem{Tr75}L.\ R.\ G.\ Treloar, {\it The Physics of Rubber Elasticity}, 
3rd ed., Oxford University Press, London (1975), Chapter VI. 
\bibitem{Wa72}H.R. Warner, Ind. Eng. Chem. Fundam. {\bf 11}, 169 (1972).
\bibitem{Co91}A. Cohen, Rheol. Acta {\bf 30}, 270 (1991).
\bibitem{HO97}M.\ Herrchen, H.\ C.\ \"{O}ttinger, J.\ 
Non-Newtonian Fluid Mech. {\bf 68}, 17 (1997).
\bibitem{BDJ80}R.\ B.\ Bird, P.\ J.\ Dotson, N.\ L.\ Johnson, 
J.\ Non-Newton.\ Fluid Mech.\ {\bf 7}, 213 (1980); {\bf 8}, 193 (1981) 
and {\bf 15}, 225 (1984) (errata).
\bibitem{ZKD00}V.\ B.\ Zmievski, I.\ V.\ Karlin and M.\ Deville, 
Physica A {\bf 275}, 152 (2000). 
\bibitem{Cu91}C.\ F.\ Curtiss, J. Chem. Phys. {\bf 95}, 1337 (1991).
\bibitem{Ris96}H. Risken, {\it The Fokker-Planck Equation}, Springer, 
2nd Ed., 1996.
\bibitem{Keu98}G.\ Lielens, P.\ Halin, I.\ Jaumin, R.\ Keunings, 
V.\ Legat, J.\ Non-Newton.\ Fluid Mech.\ {\bf 76}, 249 (1998).
\bibitem{Keu97}R.\ Keunings, J.\ Non-Newton.\ Fluid Mech.\  {\bf 68}, 
85 (1997).
\bibitem{Ju96}G. Junker, {\it Supersymmetric Methods in Quantum and 
Statistical Physics}, Springer 1996.
\bibitem{Wi}E. Witten, Nucl. Phys. {\bf B188}, 513 (1981); E. Witten, 
Nucl. Phys. {\bf B202}, 253 (1982).
\bibitem{Ge83}G.\'E. Gendenshte\^{\i}n, JETP Lett. {\bf 38}, 356 
(1983).


\end{references}
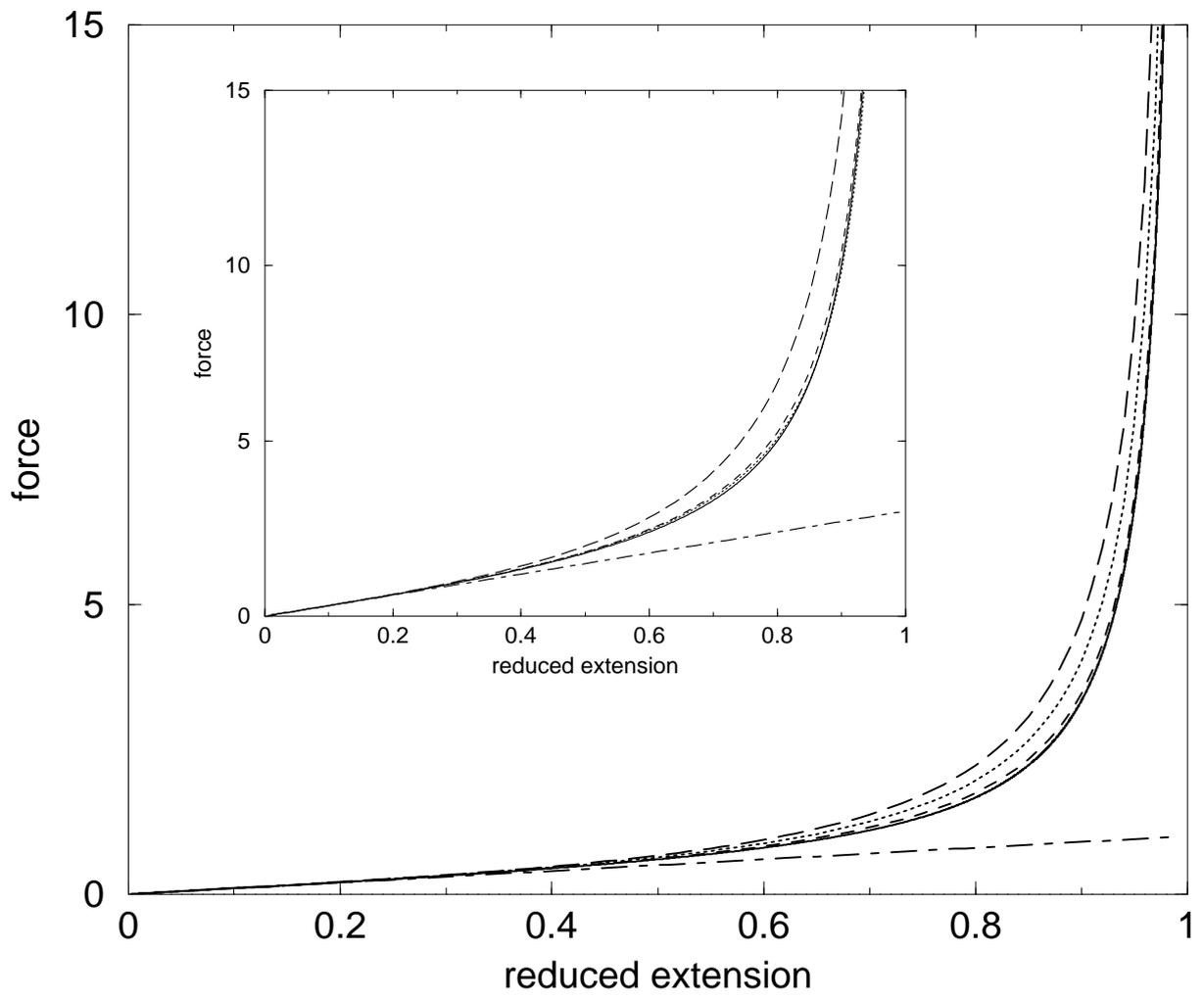
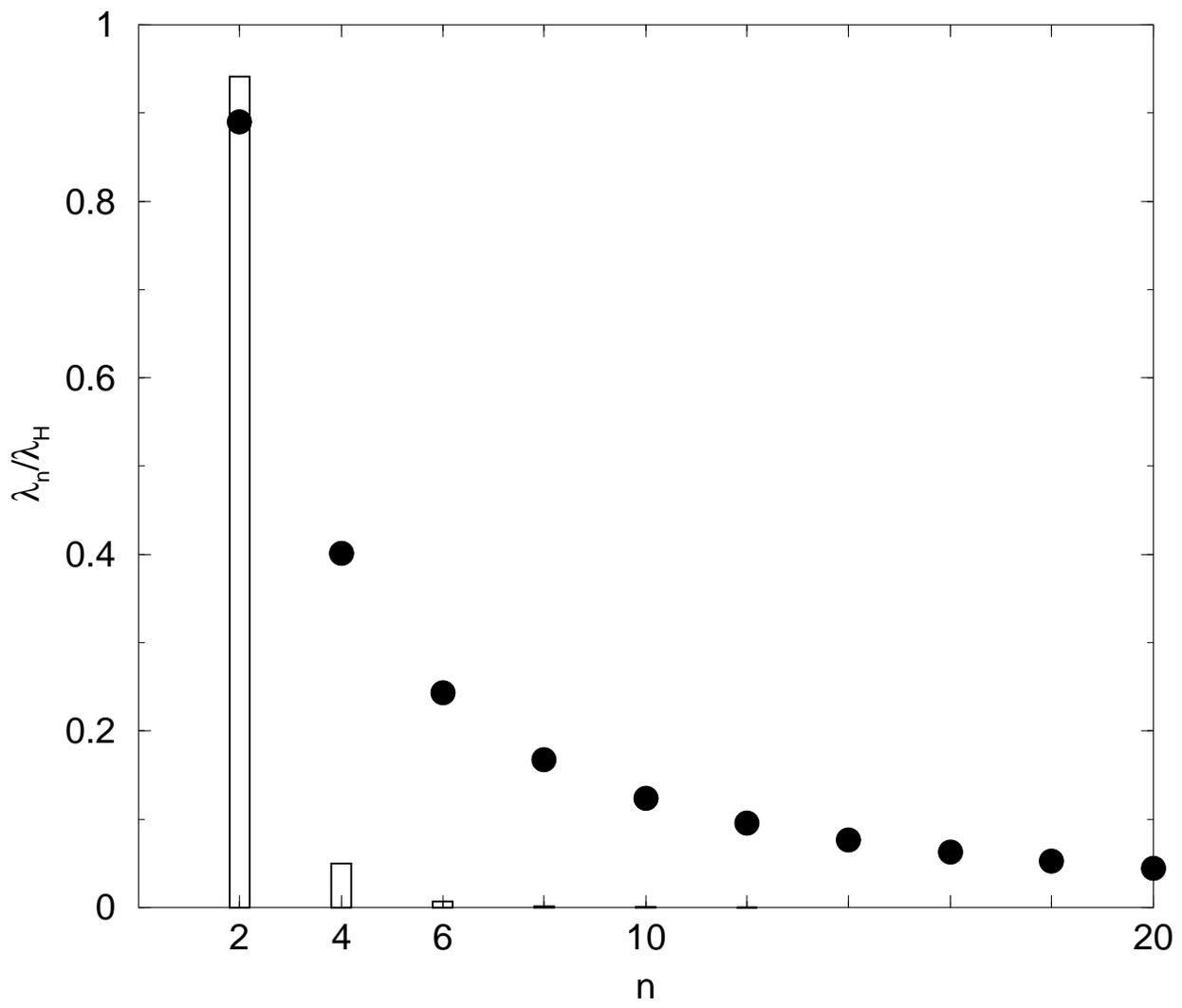
\end{document}